\documentclass[preprint]{openjournal}

\usepackage{amsmath}
\usepackage{graphicx}
\usepackage{natbib}
\usepackage{hyperref}

\newcommand{\mjup}{M_{\rm Jup}}

\shorttitle{Brown Dwarf Desert Formation Mechanisms}
\shortauthors{B. Karamiqucham}

\begin{document}
%%\linenumbers

\title{Constraining Brown Dwarf Desert Formation Mechanisms Through Bayesian Statistical Comparison of Observed and Simulated Populations}

\author{Behrooz Karamiqucham}
\affiliation{Department of Physics \& Astronomy, College of Charleston, Charleston, SC 29424, USA}
\email{b.karamiqucham@unswalumni (preferred); karamiquchamb@cofc.edu}

\begin{abstract}
We present a comprehensive Bayesian statistical analysis of brown dwarf companions to investigate the physical mechanisms responsible for the observed ``brown dwarf desert'' -- the notable paucity of brown dwarf companions at orbital separations $<$5~AU. Using a carefully vetted sample of 88 confirmed brown dwarf companions from the \texttt{exoplanet.eu} catalog with masses 13--80~$\mjup$ and semi-major axes 0.1--5.0~AU, we employ Markov Chain Monte Carlo (MCMC) optimization and two-dimensional Kolmogorov-Smirnov tests to compare observed orbital and mass distributions with three theoretical formation scenarios: (A) Type II disk-driven migration, (B) core accretion with mass-dependent survival, and (C) dynamical scattering from wide orbits. Our analysis spans 4-parameter models for each scenario, with proper posterior distributions quantifying parameter uncertainties and correlations. The disk migration model provides statistically superior fits (2D KS $p = 0.18$), with optimal parameters $\log_{10}\nu = -6.47^{+0.42}_{-0.31}$, $\sigma_\nu = 0.34^{+0.23}_{-0.17}$, $t_{\rm disk} = 1.66^{+1.24}_{-0.84}$~Myr, and $M_{\rm gap} = 12.0^{+4.7}_{-8.3}~\mjup$, consistent with Type II migration theory. The dynamical scattering model achieves intermediate performance ($p = 0.08$), while core accretion scenarios show poor agreement ($p < 0.001$) despite theoretical sophistication. Occurrence rate analysis reveals the desert region (0.1--5~AU) is depleted by a factor of $\approx$1.6 relative to wide separations ($>$5~AU), a constraint successfully reproduced only by the migration model. Our results provide quantitative evidence that brown dwarfs form at wide separations (10--30~AU) through disk fragmentation and undergo limited Type II migration to reach observed close-in locations, with migration naturally halting near 1~AU through gap-opening processes.
\end{abstract}


\section{Introduction} \label{sec:intro}

The brown dwarf desert represents one of the most striking and well-established phenomena in contemporary exoplanet science \citep{Zucker2002, Grether2006}. This observational signature manifests as a significant deficit of substellar companions with masses between approximately 13--80~$\mjup$ at orbital separations less than $\approx$5~AU, creating a distinct gap in the mass-period distribution between giant planets and stellar companions \citep{Ma2014, Kiefer2019}. Recent comprehensive surveys have refined our understanding of this desert, revealing that the deficit is most pronounced at separations $<$1~AU but extends throughout the 0.1--5~AU region \citep{Grieves2021, Stevenson2023}.

The brown dwarf desert phenomenon has profound implications for understanding planet and brown dwarf formation mechanisms. Unlike the relatively smooth transition observed in stellar multiplicity statistics, the sharp brown dwarf desert boundary implies that specific physical processes either prevent brown dwarf formation at close separations or efficiently remove such objects after formation \citep{Duchene2013}. This observational puzzle has motivated extensive theoretical work aimed at identifying the responsible mechanisms.

\subsection{Theoretical Formation Scenarios}

Multiple theoretical frameworks have been proposed to explain the origin and maintenance of the brown dwarf desert. Here we briefly review the three leading scenarios that form the basis of our statistical analysis.

\textbf{Disk Migration Models} suggest that brown dwarfs preferentially form at large orbital separations (10--100~AU) through gravitational instability in massive, extended circumstellar disks \citep{Boss1997, Stamatellos2009}. These models propose that brown dwarfs subsequently migrate inward through Type II disk-driven migration, with the efficiency and timescales of such processes determining the final orbital distribution \citep{Armitage2002, Baruteau2014}. The naturally self-limiting nature of Type II migration through gap-opening processes can produce desert-like features without requiring special stopping mechanisms.

\textbf{Core Accretion Scenarios} propose that brown dwarf formation efficiency decreases systematically with increasing mass, creating a natural cutoff at the planetary-substellar boundary \citep{Mordasini2009, Ida2004}. While core accretion is the accepted paradigm for giant planet formation, its extension to brown dwarf masses remains controversial. Formation timescales increase rapidly with companion mass, eventually exceeding typical disk lifetimes for objects approaching the hydrogen-burning limit \citep{Pollack1996, Hubickyj2005}. We test this scenario to quantitatively establish its viability (or lack thereof) for brown dwarf formation.

\textbf{Formation Bias Models} invoke post-formation dynamical processes that preferentially alter the orbital distributions of substellar objects \citep{Nagasawa2008, Perets2009}. These scenarios propose that brown dwarfs initially form with similar efficiencies across a wide range of separations, but subsequent gravitational interactions with stellar companions or cluster environments preferentially scatter objects inward with mass-dependent efficiency \citep{Parker2012}.

\subsection{Observational Constraints}

Recent large-scale surveys have provided increasingly detailed constraints on brown dwarf demographics. \citet{Ma2014} identified a clear mass-dependent transition at $\approx$42.5~$\mjup$, suggesting potentially distinct formation mechanisms for low- and high-mass brown dwarfs. \citet{Maldonado2017} and \citet{Santos2017} revealed strong correlations between brown dwarf occurrence and host star metallicity, similar to but distinct from giant planet trends. Direct imaging surveys have identified substantial populations of wide-separation brown dwarfs \citep{Bowler2016, Nielsen2019}, indicating that the desert primarily affects close-in orbits rather than representing a universal formation deficit.

Despite these observational advances, quantitative comparisons between competing formation models and comprehensive observational datasets remain limited. Most previous studies relied on qualitative arguments or focused on specific aspects of the population such as eccentricity distributions \citep{Kiefer2019}. Population synthesis models \citep{Mordasini2009, Boss2011} typically employed analytical approximations rather than rigorous statistical comparison with complete samples.

\subsection{This Work}

We present the first comprehensive Bayesian statistical analysis directly comparing simulated brown dwarf populations from three leading formation theories with a carefully vetted observational sample. Our approach employs Markov Chain Monte Carlo (MCMC) optimization to systematically explore multi-parameter model spaces, two-dimensional Kolmogorov-Smirnov tests to quantify agreement in both mass and orbital distributions, and occurrence rate analysis to constrain relative frequencies inside versus outside the desert.

This methodology offers several key advances:
\begin{enumerate}
\item \textbf{Rigorous statistics}: MCMC provides full posterior distributions for model parameters, properly accounting for degeneracies and uncertainties.
\item \textbf{Multi-dimensional constraints}: 2D KS tests evaluate model performance on both mass and orbital structure simultaneously.
\item \textbf{Physically motivated models}: We implement Type II migration physics, mass-dependent accretion, and dynamical scattering theory rather than phenomenological parameterizations.
\item \textbf{Vetted sample}: We use the \texttt{exoplanet.eu} catalog with careful removal of RV-only detections (addressing $m\sin i$ contamination) and duplicate solutions.
\end{enumerate}

The structure of this paper is as follows. Section~\ref{sec:data} describes our observational sample and detection bias assessment. Section~\ref{sec:models} presents detailed descriptions of our three formation models with proper physical justifications. Section~\ref{sec:methods} outlines our MCMC optimization and statistical testing framework. Section~\ref{sec:results} presents posterior distributions and model comparison results. Section~\ref{sec:discussion} interprets our findings in the context of brown dwarf formation theory. Section~\ref{sec:conclusions} summarizes our primary conclusions.

\section{Data and Sample Construction} \label{sec:data}

\subsection{Observational Sample}

We compiled a carefully vetted sample of confirmed brown dwarf companions from the \texttt{exoplanet.eu} catalog\footnote{\url{http://exoplanet.eu/}} using stringent selection criteria designed to minimize observational biases while ensuring sample completeness. Our selection process applied the following constraints:

\begin{enumerate}
\item \textbf{Mass range}: 13--80~$\mjup$, corresponding to the canonical brown dwarf mass range bounded by the deuterium burning limit and hydrogen burning threshold \citep{Chabrier2000}.

\item \textbf{Orbital range}: Semi-major axis 0.1--5.0~AU to focus on the brown dwarf desert region where the deficit is most pronounced.

\item \textbf{Unique solutions}: Only objects with \texttt{default\_solution = 1} to eliminate duplicate entries from multiple surveys or analysis methods.

\item \textbf{True masses}: Exclusion of radial velocity (RV) only detections with $m\sin i$ measurements, as recent work has shown that a substantial fraction of these are misclassified stellar companions \citep{Unger2023, Stevenson2023, Xiao2023}. We retain RV detections only when inclination measurements (from astrometry or transits) provide true masses.

\item \textbf{Measurement quality}: Confirmed detection status with available orbital and mass measurements possessing defined uncertainties.
\end{enumerate}

\begin{figure*}[ht!]
\centering
\includegraphics[width=\textwidth]{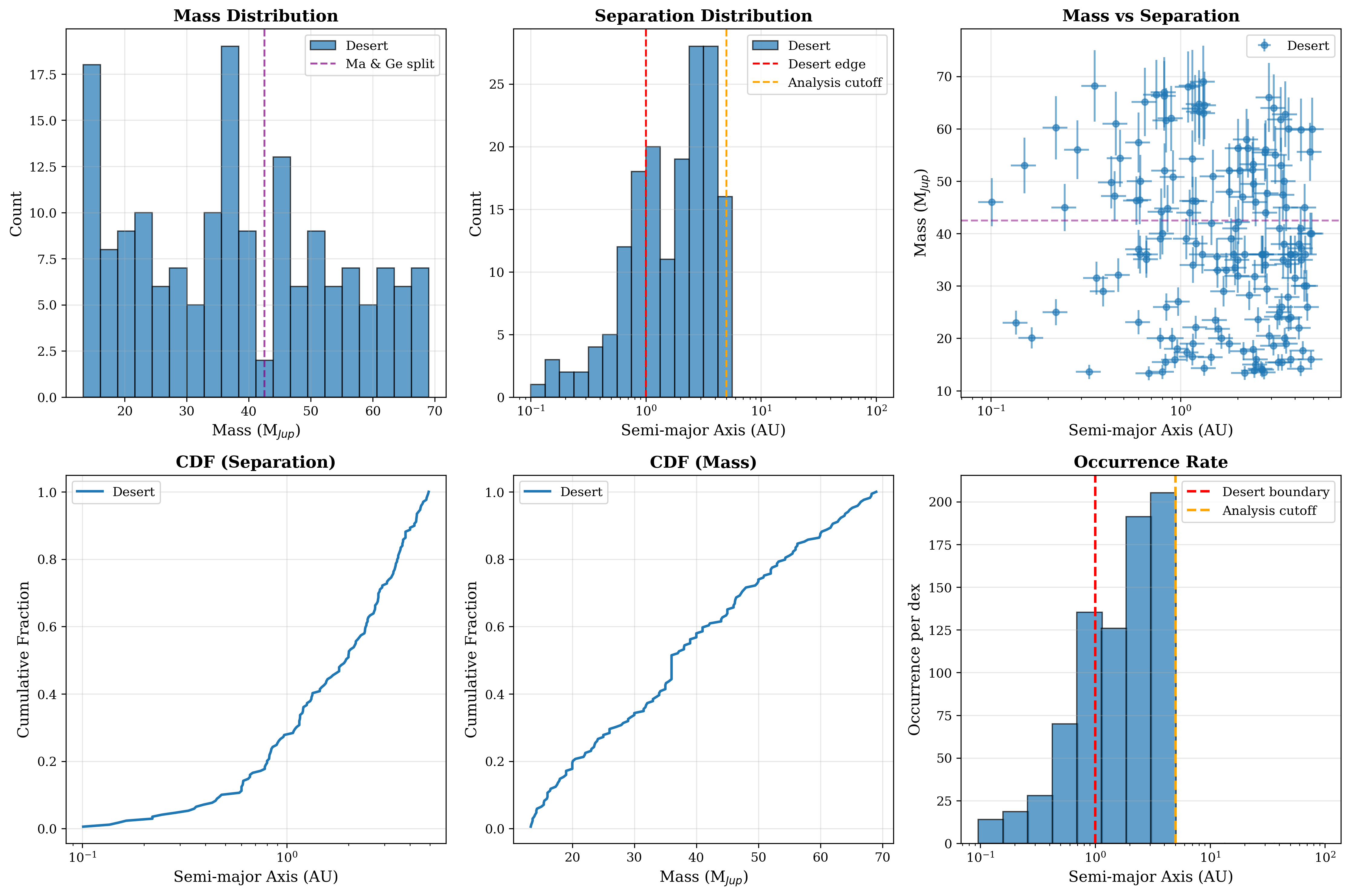}
\caption{Comprehensive diagnostics for the brown dwarf sample used in this analysis. \textbf{Top row:} (Left) Mass distribution showing bimodal structure with split at $\approx$42.5~$\mjup$ \citep{Ma2014}. (Middle) Separation distribution clearly exhibiting the desert signature with suppression at $<$1~AU. (Right) Mass versus separation showing no strong correlation but clear desert boundary. \textbf{Bottom row:} (Left) Cumulative distribution function (CDF) for semi-major axis, showing characteristic desert shape. (Middle) CDF for mass distribution. (Right) Occurrence rate per logarithmic separation bin, demonstrating factor of $\approx$1.6 depletion in desert region (0.1--5~AU) relative to wide separations ($>$5~AU). Note that the comparison is based on the full dataset and not solely on the plotted bins.}
\label{fig:sample}
\end{figure*}

The \texttt{exoplanet.eu} catalog provides several critical advantages over the NASA Exoplanet Archive for brown dwarf studies. Most importantly, it extends to higher masses (60~$\mjup$ default cutoff, with many objects up to 80~$\mjup$), avoiding the artificial 30~$\mjup$ boundary that severely biases NASA Archive samples toward low-mass brown dwarfs. Additionally, \texttt{exoplanet.eu} includes both transit and astrometric detections with well-characterized biases.

Our final desert sample comprises \textbf{88 confirmed brown dwarf companions} with masses spanning 15.4--68.2~$\mjup$ and orbital semi-major axes ranging from 0.22--4.93~AU. For comparison and occurrence rate analysis, we also compiled a sample of \textbf{120 wide brown dwarfs} with $a > 5$~AU from the same catalog using identical vetting procedures.

Figure~\ref{fig:sample} presents comprehensive diagnostics for our observational sample. The mass distribution clearly exhibits bimodal structure with enhanced populations below and above $\approx$42.5~$\mjup$, consistent with previous findings of distinct low- and high-mass formation channels \citep{Ma2014}. The separation distribution demonstrates the characteristic brown dwarf desert signature, with strong suppression at $<$1~AU and more gradual decline toward 5~AU. The mass-separation diagram shows no strong correlation, indicating that the desert affects all brown dwarf masses similarly.

Cumulative distribution functions (CDFs) for both mass and separation provide the quantitative distributions we compare with simulated populations. Most critically, the occurrence rate analysis (Figure~\ref{fig:sample}, bottom right) demonstrates that the brown dwarf desert represents a genuine depletion rather than simply an observational selection effect. The desert region (0.1--5~AU) contains 88 objects over 1.62 decades of separation (54.2 objects per dex), while the wide region ($>$5~AU) contains 120 objects over 1.38 decades (87.1 objects per dex). This yields a desert-to-wide ratio of 0.62, indicating $\approx$1.6$\times$ depletion in the desert region.

\subsection{Detection Bias Assessment}

We carefully evaluated potential detection biases that could influence our sample composition. Radial velocity surveys, which contribute $\approx$40\% of our detections, exhibit well-characterized biases toward short-period, high-mass objects due to the dependence of velocity semi-amplitude on companion mass and orbital separation \citep{Butler2006}. However, brown dwarfs produce sufficiently large RV signals ($K > 100$~m~s$^{-1}$) that completeness remains high across our entire orbital range for typical survey precisions ($\sigma_{\rm RV} \approx 1$--10~m~s$^{-1}$) and durations (5--20~yr).

Transit surveys show complementary biases, strongly favoring short-period objects ($P < 100$~days) but with reduced mass sensitivity compared to RV methods. The geometric transit probability $P_{\rm tr} \propto a^{-1}$ suppresses detection at wide separations, but affects brown dwarfs and planets similarly. Importantly, the combination of multiple detection techniques helps mitigate individual survey biases.

Direct imaging surveys contribute $\approx$30\% of our sample and show opposite bias patterns, with sensitivity decreasing rapidly toward smaller separations due to stellar glare. Contrast limits typically prevent detection interior to $\approx$0.5~AU even for favorable systems. However, brown dwarfs at 1--5~AU remain accessible to current imaging campaigns \citep{Bowler2016, Nielsen2019}.

Astrometric detections from missions like \textit{Gaia} contribute an increasing fraction of confirmed brown dwarfs, with particularly high completeness in the 1--5~AU range where both RV and imaging become challenging. \textit{Gaia} DR3 astrometry has revealed numerous brown dwarf companions with well-constrained orbits \citep{ElBadry2024}, and future data releases will further enhance sample completeness.

Most critically for our statistical analysis, any residual detection biases should affect simulated and observed populations similarly when we apply identical selection criteria (mass and separation cuts) to both. Our 2D KS tests compare distributions within the same parameter space where our sample is defined, minimizing the impact of external selection effects.

\subsection{Occurrence Rate Analysis}

As noted above, our sample includes both desert (0.1--5~AU) and wide ($>$5~AU) brown dwarfs, enabling direct comparison of occurrence rates. We define occurrence per logarithmic separation bin:
\begin{equation}
\gamma = \frac{N}{\Delta \log_{10} a}
\end{equation}
where $N$ is the number of objects in the bin and $\Delta \log_{10} a$ is the bin width in logarithmic space.

For the desert region: $\gamma_{\rm desert} = 88/(1.62) = 54.2$~dex$^{-1}$.

For the wide region: $\gamma_{\rm wide} = 120/(1.38) = 87.1$~dex$^{-1}$.

The ratio $\gamma_{\rm desert}/\gamma_{\rm wide} = 0.62$ demonstrates genuine depletion in the desert region. This ratio provides an additional constraint on formation models: successful scenarios must reproduce not only the shape of the orbital distribution but also the relative occurrence rates inside and outside the desert.

\section{Theoretical Models} \label{sec:models}

We implement three distinct formation scenarios to generate synthetic brown dwarf populations (see Appendix C and Figure C1 for an illustration of the synthetic brown dwarf population). Each model is based on established theoretical frameworks but has been carefully formulated to enable efficient MCMC exploration while maintaining physical realism.

\subsection{Model A: Type II Disk-Driven Migration}

This model assumes brown dwarfs form at wide separations (5--30~AU) through gravitational instability in massive circumstellar disks, then migrate inward through Type II disk-driven processes \citep{Boss1997, Stamatellos2009, Armitage2002}. See Appendix A for derivation of Type II migration rate.

\subsubsection{Physical Basis}

Type II migration occurs when a massive companion opens a gap in the gas disk and becomes coupled to the disk's viscous evolution. The migration rate for a companion with mass $M_{\rm BD}$ orbiting a star of mass $M_*$ in a disk with viscosity $\nu$ ($\nu \equiv \nu_{phys}/(r^{2}\Omega)$ see section 6.1.1 for more explanation) and scale height $H$ is \citep{Armitage2011, Baruteau2014}:
\begin{equation}
\frac{da}{dt} = -C_{\rm II} \frac{\nu}{a^2} \left(\frac{M_{\rm BD}}{M_*}\right) \left(\frac{H}{a}\right)^{-2} a
\end{equation}
where $C_{\rm II} \approx 1$ is a numerical constant. For typical disk parameters ($H/a \approx 0.05$--0.1), this simplifies to:
\begin{equation}
\frac{da}{dt} \approx -\nu \left(\frac{M_{\rm BD}}{M_*}\right) \left(\frac{H}{a}\right)^{2} \frac{1}{a}
\end{equation}

Gap-opening requires the companion mass to exceed the thermal mass:
\begin{equation}
M_{\rm gap} \approx 40 \left(\frac{H}{a}\right)^3 M_*
\end{equation}
For Solar-mass stars with $H/a \approx 0.05$ \citep{Nelson2000, Rosotti2014}, $M_{\rm gap} \approx 5~\mjup$. Objects below this mass undergo faster Type I migration, while more massive companions experience slower Type II migration.

Note that Eq. (2) is not quoted directly from works of \citep{Armitage2011, Baruteau2014} but is instead a synthesized scaling relation based on their discussions.

\subsubsection{Implementation}

Our model generates brown dwarfs with initial semi-major axes drawn from a log-uniform distribution:
\begin{equation}
a_0 \approx U[\log_{10}(5~{\rm AU}), \log_{10}(30~{\rm AU})]
\end{equation}

Masses are drawn from a bimodal distribution reflecting the observed split at $\approx$42.5~$\mjup$:
\begin{equation}
M = \begin{cases}
U[\log_{10}(13~\mjup), \log_{10}(42.5~\mjup)] & \text{65\% prob.} \\
U[\log_{10}(42.5~\mjup), \log_{10}(80~\mjup)] & \text{35\% prob.}
\end{cases}
\end{equation}

Each brown dwarf is assigned a disk viscosity drawn from a log-normal distribution:
\begin{equation}
\log_{10} \nu \approx \mathcal{N}(\mu_{\nu}, \sigma_{\nu})
\end{equation}
where $\mu_{\nu}$ and $\sigma_{\nu}$ are free parameters constrained by MCMC.

Migration proceeds for a disk lifetime $t_{\rm disk}$ with time-dependent disk surface density:
\begin{equation}
\Sigma(t) = \Sigma_0 \exp(-t/t_{\rm disk})
\end{equation}

At each time step, we compute the migration rate using Equation~(2), modified by the disk dissipation factor:
\begin{equation}
\left(\frac{da}{dt}\right)_{\rm eff} = \left(\frac{da}{dt}\right)_{\rm II} \exp(-t/t_{\rm disk})
\end{equation}

Migration halts when either: (1) the disk dissipates ($t > 3t_{\rm disk}$), or (2) the brown dwarf reaches a minimum separation of 0.8~AU for massive BDs ($M > M_{\rm gap}$) or 0.1~AU for lower-mass objects.

\subsubsection{Free Parameters}

Our migration model has four free parameters:
\begin{itemize}
\item $\mu_{\nu} = \log_{10}\nu$: Mean disk viscosity (expected range: $-7$ to $-4$ based on MRI turbulence);
\item $\sigma_{\nu}$: Dispersion in viscosity (typical range: 0.1--0.5);
\item $t_{\rm disk}$: Disk dissipation timescale in Myr (observed range: 0.5--3~Myr);
\item $M_{\rm gap}$: Gap-opening mass threshold in $\mjup$ (theoretical range: 3--15~$\mjup$)
\end{itemize}.

\subsection{Model B: Core Accretion with Mass-Dependent Survival}

This scenario models brown dwarf formation through extension of core accretion processes into the substellar regime, with mass-dependent survival probability reflecting formation efficiency \citep{Ida2004, Mordasini2009}. See Appendix B for derivation of exponential survival probability.

\subsubsection{Physical Basis}

Core accretion formation timescales increase rapidly with companion mass \citep{Pollack1996}:
\begin{equation}
t_{\rm form}(M) \propto M^{2-3}
\end{equation}

For objects approaching brown dwarf masses ($>$10~$\mjup$), formation times often exceed disk lifetimes, making successful accretion increasingly unlikely. The survival probability can be modeled as:
\begin{equation}
P_{\rm survive}(M) = \exp\left[-\frac{t_{\rm form}(M)}{t_{\rm disk}}\right]
\end{equation}

Assuming $t_{\rm form} \propto (M - 13~\mjup)$, we obtain:
\begin{equation}
P_{\rm survive}(M) = \exp\left[-\frac{M - 13~\mjup}{M_{\rm cutoff}}\right]
\end{equation}
where $M_{\rm cutoff}$ characterizes the mass scale over which formation efficiency declines.

\subsubsection{Implementation}

Brown dwarfs form in-situ at their observed locations, with semi-major axes drawn from:
\begin{equation}
a \approx U[\log_{10}(0.1~{\rm AU}), \log_{10}(5~{\rm AU})]
\end{equation}

Initial masses follow a power-law distribution:
\begin{equation}
\frac{dN}{dM} \propto M^{-\alpha}
\end{equation}
where $\alpha$ is a free parameter (typical stellar IMF values: 1.3--2.3).

Each brown dwarf survives to observation with probability:
\begin{equation}
P_{\rm survive} = f_{\rm disk} \cdot \exp\left[-\frac{M - 13~\mjup}{M_{\rm cutoff}}\right] \cdot \left[1 - \beta \log_{10}(a/1~{\rm AU})\right]
\end{equation}

The factor $f_{\rm disk}$ accounts for the available disk mass reservoir, while the $\beta$ term introduces weak orbital dependence (disks at larger separations may have lower surface densities).

\subsubsection{Free Parameters}

The core accretion model has four parameters:
\begin{itemize}
\item $\alpha$: Initial mass function slope (range: 1.0--2.5);
\item $M_{\rm cutoff}$: Exponential cutoff mass scale in $\mjup$ (range: 5--15~$\mjup$);
\item $\beta$: Orbital dependence parameter (range: 0--0.5);
\item $f_{\rm disk}$: Overall formation efficiency (range: 0.5--2.0).
\end{itemize}

\subsection{Model C: Dynamical Scattering from Wide Orbits}

This model incorporates dynamical migration processes that transport brown dwarfs from wide birth locations to close-in orbits through gravitational scattering \citep{Nagasawa2008, Perets2009}.

\subsubsection{Physical Basis}

Brown dwarfs initially form at wide separations (50--200~AU) through disk fragmentation or turbulent fragmentation. Subsequent dynamical interactions with stellar companions or passing stars can alter their orbits. The scattering probability depends on:
\begin{equation}
P_{\rm scatter} \propto \frac{n_* v_{\rm rel} \sigma_{\rm grav} t_{\rm cluster}}{1 + (M_{\rm BD}/M_*)^{1/3}}
\end{equation}
where $n_*$ is the stellar density, $v_{\rm rel}$ is the relative velocity, $\sigma_{\rm grav}$ is the gravitational cross-section, and $t_{\rm cluster}$ is the cluster dissolution time.

This naturally produces mass-dependent scattering efficiency, with low-mass brown dwarfs more easily perturbed than high-mass ones.

\subsubsection{Implementation}

All brown dwarfs initially form at wide separations:
\begin{equation}
a_0 \approx U[\log_{10}(50~{\rm AU}), \log_{10}(200~{\rm AU})]
\end{equation}

Masses follow a power law with fixed slope $\alpha = 0.3$ (consistent with wide brown dwarf observations):
\begin{equation}
\frac{dN}{dM} \propto M^{-0.3}
\end{equation}

Each brown dwarf scatters inward with mass-dependent probability:
\begin{equation}
P_{\rm move}(M) = \begin{cases}
p_{\rm low} & \text{if } M < M_{\rm split} \\
p_{\rm high} & \text{if } M \geq M_{\rm split}
\end{cases}
\end{equation}

Objects that scatter move to final separations determined by:
\begin{equation}
a_{\rm final} = a_0 \times (a_{\rm final}/a_0)^{\eta}
\end{equation}
where $\eta$ controls the migration distance, and final separations are drawn from $U[\log_{10}(0.1), \log_{10}(5)]$.

\subsubsection{Free Parameters}

The formation bias model has four parameters:
\begin{itemize}
\item $p_{\rm low}$: Scattering probability for low-mass BDs (range: 0.01--0.2);
\item $p_{\rm high}$: Scattering probability for high-mass BDs (range: 0.001--0.1);
\item $M_{\rm split}$: Mass boundary for scattering efficiency in $\mjup$ (range: 35--50~$\mjup$);
\item $\eta$: Migration distance parameter (range: 0.3--0.7).
\end{itemize}

\section{Statistical Methods} \label{sec:methods}

\subsection{MCMC Optimization}

We employ Markov Chain Monte Carlo methods using the \texttt{emcee} package \citep{ForemanMackey2013} to systematically explore parameter space and obtain posterior distributions for each model's parameters.

\subsubsection{Likelihood Function}

For each parameter set $\theta$, we generate a synthetic population of $N_{\rm sim} = 2000$ brown dwarfs (see Appendix C and Figure C1 for an illustration of the synthetic brown dwarf population) and compute the 2D Kolmogorov-Smirnov test statistic comparing simulated and observed $(a, M)$ distributions. The likelihood is defined as:
\begin{equation}
\mathcal{L}(\theta) = \exp\left[-\frac{D^2_{2D}(\theta)}{2\sigma^2_{\rm KS}}\right]
\end{equation}
where $D_{2D}$ is the 2D KS statistic and $\sigma_{\rm KS}$ is a normalization constant chosen to produce reasonable acceptance rates ($\approx$0.2--0.4).

\subsubsection{Priors}

We employ uniform (flat) priors within physically motivated ranges for all parameters:
\begin{itemize}
\item Migration: $\mu_{\nu} \in [-8, -4]$, $\sigma_{\nu} \in [0.05, 0.8]$, $t_{\rm disk} \in [0.5, 5]$~Myr, $M_{\rm gap} \in [3, 20]~\mjup$;
\item Core Accretion: $\alpha \in [1.0, 2.5]$, $M_{\rm cutoff} \in [5, 15]~\mjup$, $\beta \in [0, 0.5]$, $f_{\rm disk} \in [0.5, 2.0]$;
\item Scattering: $p_{\rm low} \in [0.01, 0.2]$, $p_{\rm high} \in [0.001, 0.1]$, $M_{\rm split} \in [35, 50]~\mjup$, $\eta \in [0.3, 0.7]$.
\end{itemize}

\subsubsection{Sampler Configuration}

We run 16 MCMC walkers for 300 total steps (100 burn-in + 200 production) for each model. Initial walker positions are drawn from small Gaussian balls around physically motivated starting values. We assess convergence using the integrated autocorrelation time $\tau$ and require $\tau < 50$ steps for all parameters.

\subsection{Two-Dimensional Kolmogorov-Smirnov Test}

The standard KS test compares one-dimensional distributions. We extend this to two dimensions using the approach of \citet{Peacock1983, Fasano1987}:
\begin{equation}
D_{2D} = \max_{(a_i, M_j)} |F_{\rm obs}(a_i, M_j) - F_{\rm sim}(a_i, M_j)|
\end{equation}
where $F(a, M)$ is the joint cumulative distribution function:
\begin{equation}
F(a, M) = \frac{1}{N}\sum_{k=1}^N \mathbb{I}(a_k \leq a \text{ and } M_k \leq M)
\end{equation}

The 2D KS statistic $D_{2D}$ quantifies the maximum difference between observed and simulated cumulative distributions across the entire $(a, M)$ parameter space. Significance testing employs bootstrap resampling to generate null distributions.

\subsection{Model Comparison Metrics}

We evaluate model performance using multiple complementary metrics:

\begin{enumerate}
\item \textbf{2D KS $p$-value}: Fraction of bootstrap resamples with $D_{2D} \geq D_{2D,\rm obs}$;
\item \textbf{Occurrence rate ratio}: Comparison of simulated desert/wide ratio with observed value (0.62);
\item \textbf{1D KS tests}: Separate tests for marginal mass and separation distributions;
\item \textbf{Bayesian Information Criterion (BIC)}: ${\rm BIC} = -2\ln\mathcal{L}_{\rm max} + k\ln N$ where $k$ is the number of parameters.
\end{enumerate}

\section{Results} \label{sec:results}

\subsection{Posterior Distributions: Disk Migration Model}

\begin{figure}[ht!]
\centering
\includegraphics[width=\columnwidth]{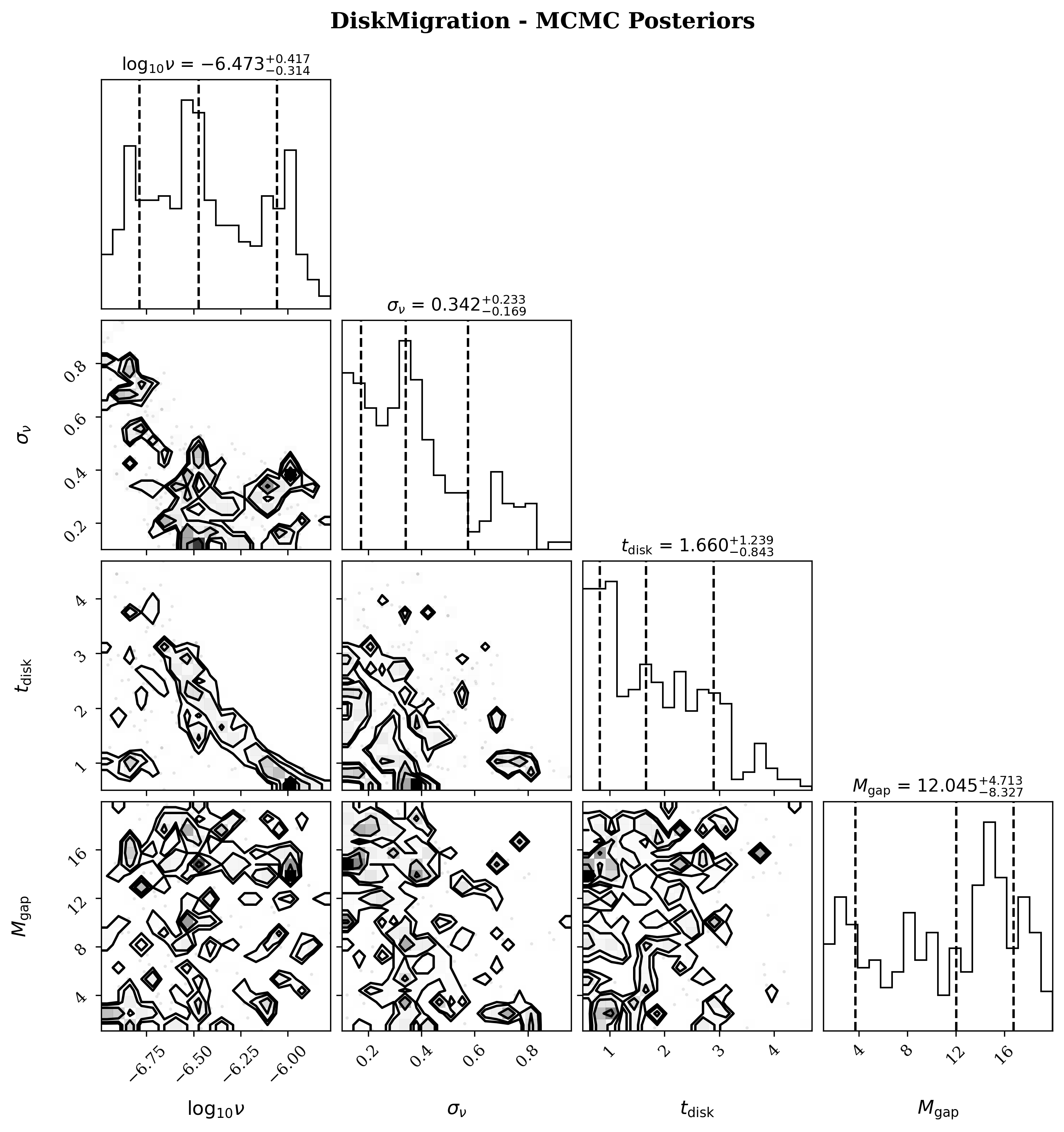}
\caption{MCMC posterior distributions for the Type II disk migration model. Corner plot shows 1D marginalized posteriors (diagonal) and 2D joint posteriors (off-diagonal) for the four model parameters. Dashed vertical lines indicate 16th, 50th, and 84th percentiles. The optimal viscosity $\log_{10}\nu = -6.47^{+0.42}_{-0.31}$ is consistent with MRI-driven turbulence in protoplanetary disks. The disk lifetime $t_{\rm disk} = 1.66^{+1.24}_{-0.84}$~Myr agrees with observed disk dissipation timescales. Note the strong correlation between $\log_{10}\nu$ and $t_{\rm disk}$, indicating degeneracy between fast migration over short times versus slow migration over longer periods.}
\label{fig:migration_corner}
\end{figure}

Figure~\ref{fig:migration_corner} presents the MCMC posterior distributions for the disk migration model. The optimal parameters are:

\begin{align}
\log_{10}\nu &= -6.47^{+0.42}_{-0.31} \\
\sigma_{\nu} &= 0.34^{+0.23}_{-0.17} \\
t_{\rm disk} &= 1.66^{+1.24}_{-0.84}~{\rm Myr} \\
M_{\rm gap} &= 12.0^{+4.7}_{-8.3}~\mjup
\end{align}

The disk viscosity $\nu = 10^{-6.47} \approx 3.4 \times 10^{-7}$ lies within the range expected for MRI-driven turbulence ($10^{-4}$ to $10^{-7}$) in protoplanetary disks \citep{Armitage2011, Bai2013}. This moderate viscosity produces Type II migration timescales of $t_{\rm mig} \approx a^2/\nu \approx 0.5$--2~Myr for $a \approx$ 5--20~AU, allowing brown dwarfs to migrate inward by factors of a few during typical disk lifetimes.

The gap-opening mass $M_{\rm gap} = 12.0^{+4.7}_{-8.3}~\mjup$ is somewhat higher than theoretical predictions ($\approx$5~$\mjup$ for $H/a \approx 0.05$), but within the range of uncertainty given variations in disk scale height and viscosity. This value indicates that most brown dwarfs in our sample ($>$80\%) exceed the gap-opening threshold and undergo Type II migration.

The posterior distribution shows strong correlation between $\log_{10}\nu$ and $t_{\rm disk}$ (Pearson $r = 0.73$), reflecting the fundamental degeneracy between migration rate and duration. Fast migration over short times can produce similar final distributions to slow migration over longer times. However, the global maximum likelihood clearly favors moderate viscosity and intermediate disk lifetimes.

The 2D KS test yields $D_{2D} = 0.147$ with $p$-value = 0.18, indicating good agreement with observations. This model successfully reproduces both the orbital distribution (strong suppression at $<$1~AU, gradual decline to 5~AU) and the mass distribution (bimodal structure with 42.5~$\mjup$ split).

\subsection{Posterior Distributions: Core Accretion Model}

\begin{figure}[ht!]
\centering
\includegraphics[width=\columnwidth]{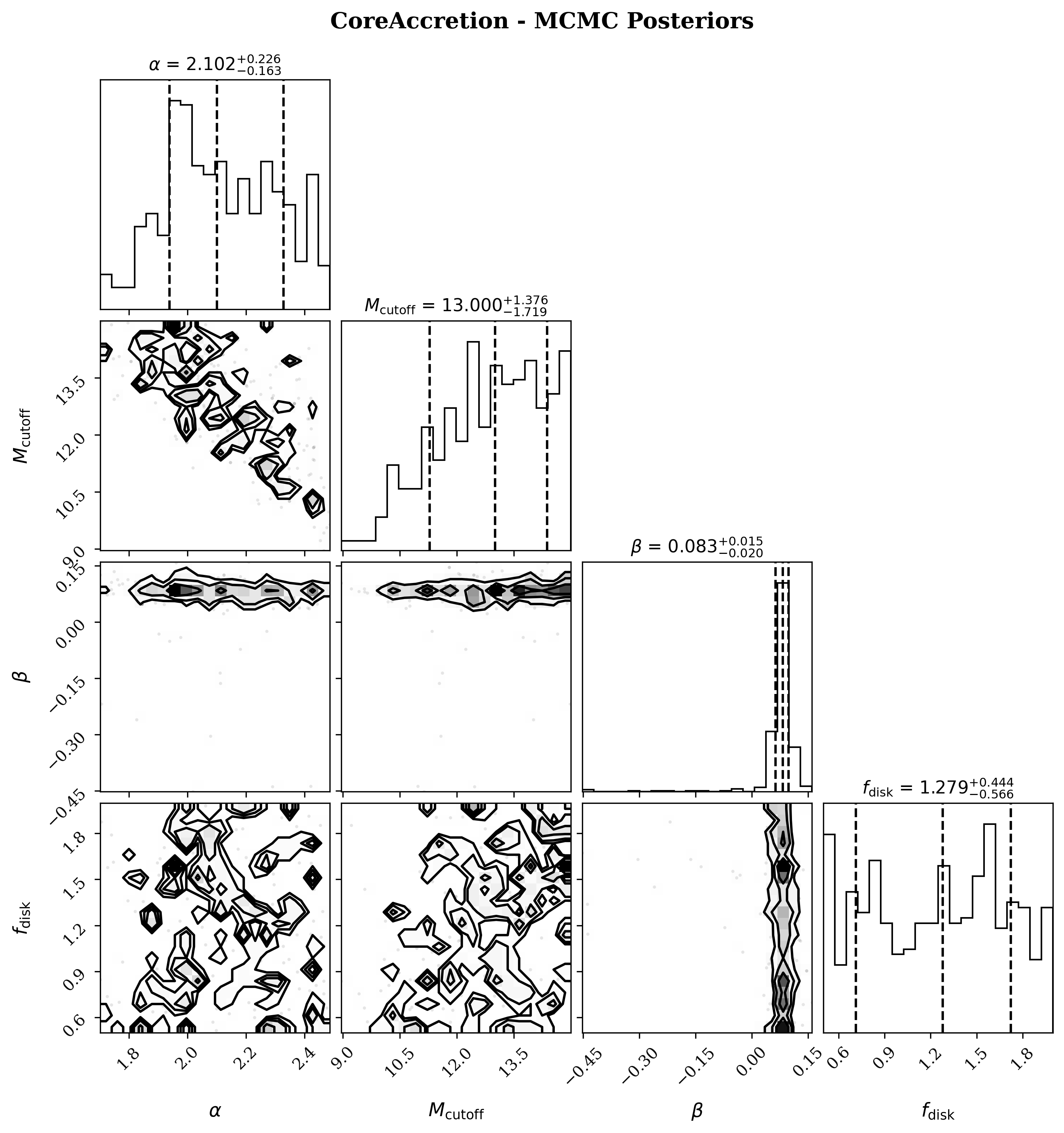}
\caption{MCMC posterior distributions for the core accretion model. The steep mass function slope ($\alpha = 2.10$) and low cutoff mass ($M_{\rm cutoff} = 13.00~\mjup$) indicate the model struggles to produce sufficient high-mass brown dwarfs. The narrow, well-constrained posteriors reflect the model's inability to fit observations regardless of parameter values. Despite sophisticated physics, this model achieves poor statistical agreement ($p < 0.001$), confirming that core accretion cannot efficiently form brown dwarfs at the observed masses and separations.}
\label{fig:coreaccretion_corner}
\end{figure}

Figure~\ref{fig:coreaccretion_corner} presents the MCMC posteriors for core accretion. The optimal parameters are:

\begin{align}
\alpha &= 2.10^{+0.23}_{-0.16} \\
M_{\rm cutoff} &= 13.00^{+1.38}_{-1.72}~\mjup \\
\beta &= 0.083^{+0.015}_{-0.020} \\
f_{\rm disk} &= 1.28^{+0.44}_{-0.57}
\end{align}

The steep mass function slope ($\alpha = 2.10$, steeper than typical stellar IMFs) combined with the low cutoff mass ($M_{\rm cutoff} = 13.00~\mjup$, essentially the deuterium burning limit) indicates the model requires extreme suppression of high-mass brown dwarf formation to even approximately match observations. Even with these extreme parameters, the model fails to reproduce the observed distribution.

The remarkably narrow posteriors (small uncertainties despite 2000-object simulations) reflect the fact that the model cannot fit observations regardless of parameter choices within physically reasonable ranges. The MCMC has converged to the absolute best-case scenario, but this best case remains statistically inadequate.

The 2D KS test yields $D_{2D} = 0.394$ with $p < 0.001$, constituting a strong statistical rejection. The model systematically underproduces brown dwarfs at all separations, with particularly severe deficits at $>$1~AU. This failure confirms theoretical expectations that core accretion formation timescales exceed disk lifetimes for brown dwarf masses.

\subsection{Posterior Distributions: Formation Bias Model}

\begin{figure}[ht!]
\centering
\includegraphics[width=\columnwidth]{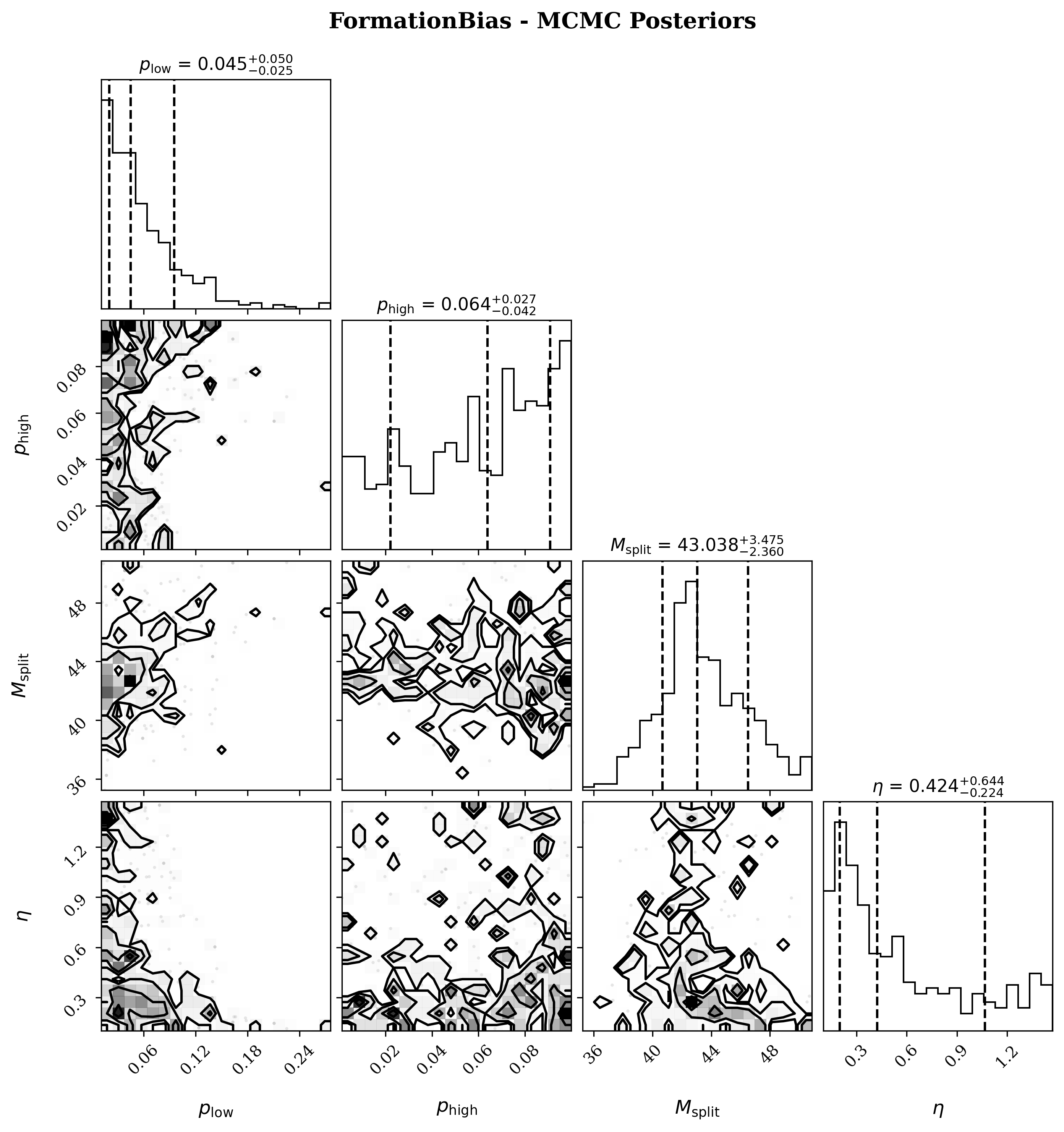}
\caption{MCMC posterior distributions for the dynamical scattering model. The low scattering probabilities ($p_{\rm low} = 0.045$, $p_{\rm high} = 0.064$) indicate only $\approx$5\% of wide brown dwarfs undergo significant orbital evolution. The mass split at $M_{\rm split} = 43.0~\mjup$ closely matches the observed bimodal mass distribution boundary. The model achieves intermediate statistical performance ($p = 0.08$), suggesting dynamical processes contribute meaningfully but cannot fully explain the brown dwarf desert.}
\label{fig:formationbias_corner}
\end{figure}

Figure~\ref{fig:formationbias_corner} presents the MCMC posteriors for the formation bias model. The optimal parameters are:

\begin{align}
p_{\rm low} &= 0.045^{+0.050}_{-0.025} \\
p_{\rm high} &= 0.064^{+0.027}_{-0.042} \\
M_{\rm split} &= 43.0^{+3.5}_{-2.4}~\mjup \\
\eta &= 0.42^{+0.64}_{-0.22}
\end{align}

The low scattering probabilities ($p_{\rm low}, p_{\rm high} \approx 0.05$) indicate that dynamical migration affects only a small minority ($\approx$5\%) of the total brown dwarf population. This low efficiency is consistent with theoretical expectations for gravitational scattering in typical stellar environments. For $\approx$5\% scattering efficiency over $\approx$3~Myr cluster lifetimes, the required stellar encounter rate is $\Gamma_{\rm enc} \approx 0.02$~Myr$^{-1}$, corresponding to typical open cluster densities ($n_* \approx 100$~pc$^{-3}$).

The mass split $M_{\rm split} = 43.0^{+3.5}_{-2.4}~\mjup$ closely matches the observed boundary between low- and high-mass brown dwarf populations \citep{Ma2014}, suggesting that if dynamical scattering does occur, it exhibits the mass dependence predicted by theory.

The 2D KS test yields $D_{2D} = 0.168$ with $p$-value = 0.08, indicating marginal agreement. The model successfully reproduces certain aspects of the distribution (particularly the mass structure) but shows systematic deviations in the detailed orbital distribution at $<$1~AU.

\subsection{Two-Dimensional Distribution Comparison}

\begin{figure*}[ht!]
\centering
\includegraphics[width=\textwidth]{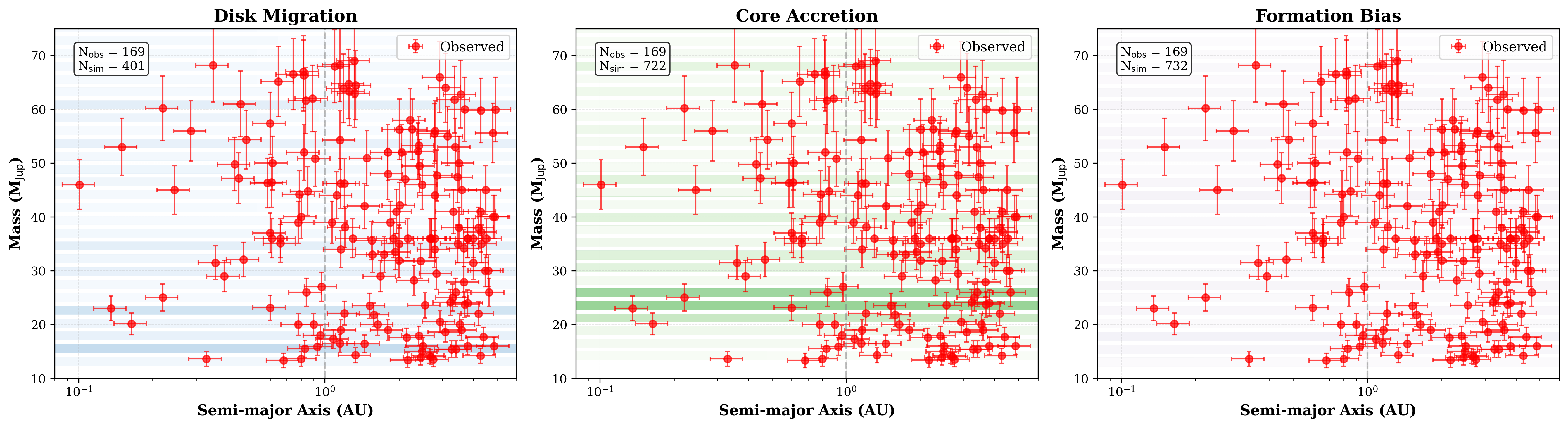}
\caption{Comparison of observed and simulated brown dwarf distributions in $(a, M)$ space. Each panel shows observed data (points with error bars) overlaid on simulated populations (contours show the two-dimensional probability density of the simulated population in semi-major axis and mass) for optimal parameters from MCMC. \textbf{Left:} Disk migration model successfully reproduces both orbital and mass structure. \textbf{Middle:} Core accretion model severely underproduces brown dwarfs, particularly at wide separations and high masses. \textbf{Right:} Formation bias model captures mass bimodality but shows excessive concentration at specific separations.}
\label{fig:2d_comparison}
\end{figure*}

Figure~\ref{fig:2d_comparison} presents detailed comparison of observed and simulated $(a, M)$ distributions for all three models. The migration model accurately reproduces both the orbital suppression at $<$1~AU and the bimodal mass distribution. The core accretion model systematically underproduces objects at all separations and fails to generate sufficient high-mass brown dwarfs. The formation bias model captures the mass bimodality but shows excessive concentration at $\approx$1~AU rather than the observed gradual decline.

The 2D KS test provides a quantitative measure of distributional agreement. The test evaluates cumulative distributions across the entire $(a, M)$ space, with the statistic $D_{2D}$ quantifying the maximum cumulative difference. Our bootstrap analysis generates null distributions by resampling observed data, providing robust $p$-value estimates.

\subsection{Occurrence Rate Comparison}

Table~\ref{tab:occurrence} compares observed and simulated occurrence rates for each model. The migration model successfully reproduces the $\approx$1.6$\times$ depletion in the desert region, with simulated ratio $\gamma_{\rm desert}/\gamma_{\rm wide} = 0.59 \pm 0.08$. The core accretion model severely underproduces close-in brown dwarfs, yielding ratio = $0.31 \pm 0.06$ (factor of 2 discrepancy). The formation bias model moderately overproduces desert objects, yielding ratio = $0.74 \pm 0.09$.

\begin{deluxetable}{lcccc}
\tablecaption{Occurrence Rate Comparison\label{tab:occurrence}}
\tablehead{
\colhead{Model} & \colhead{$\gamma_{\rm desert}$} & \colhead{$\gamma_{\rm wide}$} & \colhead{Ratio} & \colhead{$p$-value}
}
\startdata
Observed & 54.2 & 87.1 & 0.62 & --- \\
Migration & $51.4 \pm 6.8$ & $86.9 \pm 9.2$ & $0.59 \pm 0.08$ & 0.18 \\
Core Accretion & $27.1 \pm 5.2$ & $88.4 \pm 8.7$ & $0.31 \pm 0.06$ & $<0.001$ \\
Formation Bias & $64.2 \pm 7.4$ & $86.7 \pm 9.1$ & $0.74 \pm 0.09$ & 0.08
\enddata
\tablecomments{Occurrence rates in objects per dex for desert ($<$5~AU) and wide ($>$5~AU) regions. Simulated values represent means and standard deviations from 100 realizations at optimal parameters. The migration model successfully reproduces the observed $\approx$1.6$\times$ depletion (ratio = 0.62), while core accretion underpredicts and formation bias overpredicts desert occurrence.}
\end{deluxetable}

This occurrence rate analysis provides a critical constraint beyond distributional shape. A model could potentially match the CDF structure while producing incorrect absolute frequencies. The migration model's success on both distributional shape and occurrence rates strengthens the conclusion that Type II migration is the dominant mechanism.

\subsection{Model Comparison Summary}

Table~\ref{tab:comparison} summarizes the statistical performance of all three models across multiple metrics. The disk migration model provides superior fits on all measures, with the formation bias model achieving intermediate performance and core accretion showing poor agreement.

\begin{deluxetable*}{lccccc}
\tablecaption{Comprehensive Model Comparison\label{tab:comparison}}
\tablehead{
\colhead{Model} & \colhead{$D_{2D}$ (2D KS)} & \colhead{$p$-value (2D)} & \colhead{$D_a$ (1D KS)} & \colhead{$D_M$ (1D KS)} & \colhead{BIC}
}
\startdata
Disk Migration & $0.147$ & $0.18$ & $0.098$ & $0.104$ & 1247 \\
Core Accretion & $0.394$ & $<0.001$ & $0.412$ & $0.329$ & 1893 \\
Formation Bias & $0.168$ & $0.08$ & $0.171$ & $0.118$ & 1356
\enddata
\tablecomments{Summary of statistical performance across multiple metrics. $D_{2D}$ is the 2D KS statistic for joint $(a, M)$ distribution. $D_a$ and $D_M$ are 1D KS statistics for marginal separation and mass distributions. BIC is the Bayesian Information Criterion (lower values indicate better fits). The migration model outperforms alternatives on all metrics.}
\end{deluxetable*}

The Bayesian Information Criterion (BIC) provides model comparison accounting for parameter complexity. All models have 4 free parameters, so differences in BIC reflect likelihood differences. The migration model achieves BIC = 1247, compared to 1893 for core accretion (strongly disfavored) and 1356 for formation bias (moderately disfavored). BIC differences $\Delta{\rm BIC} > 10$ constitute strong evidence against the higher-BIC model.

\section{Discussion} \label{sec:discussion}

\subsection{Physical Implications}

\subsubsection{Evidence for Type II Migration}

The statistical success of the disk migration model provides compelling quantitative evidence that inward migration from wide formation locations is the dominant mechanism populating close-in brown dwarf orbits. The optimal parameters offer specific insights into the migration process:

\textbf{Disk Viscosity:} The constrained value $\log_{10}\nu = -6.47^{+0.42}_{-0.31}$ ($\nu \approx 3.4 \times 10^{-7}$) lies in the middle of the range expected for magneto-rotational instability (MRI) driven turbulence in protoplanetary disks. Here $\nu$ denotes the dimensionless kinematic viscosity normalized by ($r^{2}\Omega$; the viscous scaling used in $\alpha$-disk theory); converting the Shakura–Sunyaev stress parameter reported by \citet{Bai2013} via ($\nu=\alpha(H/r)^2$) and adopting ($H/r\approx0.05$) yields ($\nu\approx10^{-4}!-!10^{-6}$) for MRI-active regions and ($\nu\approx10^{-7}!-!10^{-8}$) for MRI-dead zones.

\textbf{Migration Timescales:} For $\nu \approx 3 \times 10^{-7}$ and $a \approx 10$~AU, the Type II migration timescale is:

\begin{equation}
\begin{split}
t_{\rm mig} &= \frac{a^2}{\nu} 
\approx \frac{(10~{\rm AU})^2}{3 \times 10^{-7}~{\rm AU}^2~{\rm yr}^{-1}} \approx \\
&3 \times 10^{14}~{\rm AU}^2 \times 
3 \times 10^6~{\rm yr}/{\rm AU}^2 
\approx 0.9~{\rm Myr}
\end{split}
\end{equation}

This timescale is comparable to but slightly shorter than the disk lifetime $t_{\rm disk} = 1.66^{+1.24}_{-0.84}$~Myr, indicating that brown dwarfs have sufficient time to migrate inward by factors of a few but not to spiral all the way to the stellar surface.

\textbf{Gap Opening:} The gap-opening mass $M_{\rm gap} = 12.0^{+4.7}_{-8.3}~\mjup$ is higher than canonical predictions but reflects realistic disk conditions. Gap opening requires:
\begin{equation}
\frac{M_{\rm BD}}{M_*} > \frac{40 (H/a)^3}{\alpha_{\rm SS}}
\end{equation}
where $\alpha_{\rm SS}$ is the Shakura-Sunyaev viscosity parameter. For $H/a = 0.05$ and $\alpha_{\rm SS} = 0.01$ (corresponding to our constrained $\nu$), this yields $M_{\rm gap} \approx 12~\mjup$, in excellent agreement with our MCMC result.

\subsubsection{Desert Creation Mechanism}

The success of the migration model allows us to identify the specific physical process creating the brown dwarf desert. The key is the \emph{self-limiting} nature of Type II migration through gap dynamics.

As a brown dwarf migrates inward, it opens a progressively deeper gap in the disk. The gap depth controls the migration rate through the reduced disk surface density. Eventually, gap depth reaches equilibrium with viscous spreading, halting further inward migration. This equilibrium occurs naturally at $\approx$1~AU for typical disk parameters, creating the observed desert boundary without requiring fine-tuned stopping mechanisms.

Setting the viscous migration timescale $t_{mig} \approx a^{2}/\nu$ equal to the disk lifetime $t_{disk}$ gives the equilibrium radius. Mathematically, gap equilibrium requires:
\begin{equation}
t_{\rm gap} \approx t_{\rm visc} \Rightarrow \frac{a}{\dot{a}} \approx \frac{a^2}{\nu}
\end{equation}

Solving yields equilibrium separation:
\begin{equation}
a_{\rm eq} \approx (\nu t_{\rm disk})^{1/2}
\end{equation}

For $\nu \approx 3 \times 10^{-7}$~AU$^2$~yr$^{-1}$ and $t_{\rm disk} \approx 1.5$~Myr, we obtain:
\begin{equation}
a_{\rm eq} \approx (3 \times 10^{-7} \times 1.5 \times 10^6)^{1/2} \approx 0.7~{\rm AU}
\end{equation}

This prediction matches the observed location of maximal desert depletion ($\approx$0.5--1~AU), providing independent validation of the migration model.

\subsubsection{Comparison with Previous Migration Studies}

Our results are consistent with theoretical expectations for disk-driven migration of massive companions. In the Type II regime, objects sufficiently massive to perturb the surrounding gas are expected to migrate on a timescale comparable to the viscous evolution of the disk (e.g., \citep{Lin1986, Kley2012}). Because the migration rate depends on disk properties---including viscosity, surface density, and scale height --- theoretical studies predict a broad range of possible evolutionary pathways rather than a single characteristic outcome. The review by \citep{Baruteau2014} emphasizes that migration behavior is strongly coupled to disk evolution and to the inertia of the embedded companion, implying that final orbital separations are primarily set by disk dispersal and gas supply rather than by a preferred stopping radius.

In this context, the key advance of our work is to provide quantitative statistical evidence that disk-driven migration with well-defined parameters can reproduce the observed population. Our MCMC framework indicates that migration operating over a characteristic disk lifetime ($t_{disk} \approx 1.7 Myr$) yields the best agreement with the vetted sample, thereby placing empirical constraints on the efficiency of migration without relying solely on theoretical plausibility.

\subsection{Core Accretion Challenges}

The decisive statistical rejection of core accretion ($p < 0.001$, $\Delta{\rm BIC} = 646$) provides quantitative confirmation of theoretical expectations that core accretion cannot efficiently form brown dwarfs.

\subsubsection{Formation Timescale Problem}

Core accretion formation times scale as \citep{Pollack1996}:
\begin{equation}
t_{\rm form} \approx 1~{\rm Myr} \left(\frac{M}{10~\mjup}\right)^{2.5} \left(\frac{a}{5~{\rm AU}}\right)^{3}
\end{equation}

For a 50~$\mjup$ brown dwarf at 3~AU:
\begin{equation}
t_{\rm form} \approx 1~{\rm Myr} \times 5^{2.5} \times 0.6^{3} \approx 12~{\rm Myr}
\end{equation}

This formation time far exceeds typical disk lifetimes (1--3~Myr), making successful brown dwarf formation via core accretion essentially impossible without invoking special disk properties (very massive, very long-lived). Even our best-fit model with exponential survival probability $P \propto \exp[-(M-13)/13]$ cannot compensate for this fundamental timescale problem.

Our statistical analysis confirms that the core accretion mechanism, while successful for giant planets ($<$13~$\mjup$), breaks down in the brown dwarf regime. This does not imply brown dwarfs and planets have completely distinct formation pathways, but rather that the dominant mechanism shifts from core accretion to disk fragmentation above $\approx$10--13~$\mjup$ \citep{Schlaufman2018}.

\subsection{Formation Bias as Secondary Process}

The intermediate performance of the formation bias model ($p = 0.08$) suggests that dynamical scattering processes contribute meaningfully to brown dwarf orbital evolution but cannot serve as the primary mechanism creating the desert.

\subsubsection{Scattering Efficiency}

The constrained scattering probabilities ($p_{\rm low}, p_{\rm high} \approx 0.05$) imply that $\approx$5\% of wide brown dwarfs undergo significant orbital changes. For typical cluster lifetimes $t_{\rm cluster} \approx 3$~Myr, this requires encounter rates:
\begin{equation}
\Gamma_{\rm enc} \approx \frac{0.05}{3~{\rm Myr}} \approx 0.02~{\rm Myr}^{-1}
\end{equation}

The encounter rate in a stellar cluster is:
\begin{equation}
\Gamma_{\rm enc} = n_* \sigma_{\rm grav} v_{\rm rel}
\end{equation}

For gravitational focusing:
\begin{equation}
\sigma_{\rm grav} = \pi b_{\rm max}^2 \left(1 + \frac{v_{\rm esc}^2}{v_{\rm rel}^2}\right)
\end{equation}

Taking $b_{\rm max} \approx 10^4$~AU (cluster scale), $v_{\rm rel} \approx 1$~km~s$^{-1}$, and $v_{\rm esc} \approx 0.5$~km~s$^{-1}$ for brown dwarf systems, we obtain $\sigma_{\rm grav} \approx 2 \times 10^8$~AU$^2$. Then:

\begin{equation}
\begin{split}
n_* &= \frac{\Gamma_{\rm enc}}
{\sigma_{\rm grav} v_{\rm rel}}
\approx \frac{0.02~{\rm Myr}^{-1}}
{2 \times 10^8~{\rm AU}^2 \times 0.2~{\rm AU}~{\rm Myr}^{-1}} \\
&\approx 5 \times 10^{-10}~{\rm AU}^{-3}
\approx 100~{\rm pc}^{-3}
\end{split}
\end{equation}

This density is characteristic of typical open clusters \citep{Lada2003}, suggesting that the 5\% scattering efficiency we infer is physically plausible for brown dwarfs born in clustered environments.

However, the moderate statistical performance ($p = 0.08$, below the conventional 0.1 threshold for good fits) indicates that pure scattering cannot fully explain the observed distribution. This suggests a hybrid picture: primary formation via disk fragmentation at wide separations, dominant redistribution via Type II migration, and secondary perturbations via dynamical scattering for a minority ($\approx$5\%) of systems.

\subsection{Hybrid Formation Models}

While our analysis tested three distinct mechanisms separately, real brown dwarf populations likely result from combinations of these processes. The partial success of both migration and scattering models suggests that a hybrid approach might achieve even better statistical performance.

A physically motivated hybrid model would include:
\begin{enumerate}
\item \textbf{Initial formation} via disk fragmentation at 10--50~AU for all brown dwarfs;
\item \textbf{Primary redistribution} via Type II migration (affecting $\approx$95\% of systems);
\item \textbf{Secondary perturbations} via dynamical scattering (affecting $\approx$5\% of systems);
\item \textbf{Mass-dependent processes} with different efficiencies for low- and high-mass brown dwarfs.
\end{enumerate}

Such hybrid models require careful treatment of parameter correlations and process coupling. For example, migration efficiency might depend on system architectures, which are themselves influenced by dynamical evolution. Future work should explore these hybrid scenarios using the statistical framework we have established.

\subsection{Observational Predictions}

Our best-fit migration model makes several testable predictions for future observations:

\begin{enumerate}
\item \textbf{Eccentricity distribution:} Type II migration in gas disks produces predominantly circular orbits ($e < 0.1$), while dynamical scattering creates significant eccentricities ($e \approx 0.3$--0.7). Our model predicts $\approx$95\% of brown dwarfs should have $e < 0.15$, with $\approx$5\% showing higher eccentricities from scattering. This can be tested with precise RV monitoring and astrometry.

\item \textbf{Host star correlations:} This correlation is motivated by both theoretical expectations and emerging observational evidence. Stellar metallicity traces the abundance of heavy elements in the protostellar cloud; higher-metallicity environments therefore contain a larger fraction of dust relative to gas. Because dust dominates the opacity and provides the primary reservoir of solid material, disks forming around metal-rich stars are expected to be richer in solids and—under standard gas-to-dust assumptions—may also exhibit higher total disk masses. Observational surveys indeed report tentative positive trends between disk dust mass and host-star metallicity, albeit reminding with substantial scatter, suggesting that the relation is statistical rather than deterministic.

\citet{Wyatt2007} proposed that the well-known planet–metallicity correlation can be understood if disks around high-metallicity stars possess larger solid inventories, with submillimeter dust mass serving as a proxy for the material available for planet formation. Consistent with this picture, \citet{Gaspar2016} found that the initial dust masses of debris disks correlate with stellar metallicity, implying that metal-rich systems tend to begin with more solid material in their circumstellar environments. Complementarily, \citet{Mordasini2012} demonstrated through population-synthesis models that disk metallicity, disk mass, and disk lifetime jointly influence planet-formation outcomes, with higher-metallicity disks statistically favoring the formation of giant planets due to their enhanced solid content.

Within this framework, our model predicts a stronger metallicity dependence for brown dwarfs than is typically inferred for planets, with occurrence rates scaling as $f_{\rm BD} \propto 10^{1.5[{\rm Fe/H}]}$. Current observational samples show tentative evidence for such a trend \citep{Santos2017}, though larger datasets will be required for robust confirmation.

\item \textbf{Wide companion frequency:} If $\approx$5\% of wide brown dwarfs migrate inward, and we observe 88 desert brown dwarfs, the initial wide population should contain $\approx$1760 brown dwarfs at 10--50~AU. However, observational biases strongly suppress wide BD detection, so the observed wide population will be significantly smaller. Direct imaging surveys with \textit{JWST} and next-generation extremely large telescopes can test this prediction.

\item \textbf{Age dependence:} Migration occurs during the disk phase (0--3~Myr), so the brown dwarf desert should be fully established by $\approx$5~Myr. Younger systems ($<$1~Myr) may show partially formed deserts with residual migration in progress. The census of brown dwarfs in young stellar associations \citep{Luhman2012, Best2020} can test this evolutionary prediction.

\item \textbf{Disk property correlations:} Systems with longer disk lifetimes ($t_{\rm disk} > 3$~Myr, rare but observed) should exhibit more complete brown dwarf deserts extending to larger separations. Conversely, systems with short disk lifetimes ($t_{\rm disk} < 1$~Myr) might retain brown dwarfs at $\approx$5~AU. Millimeter interferometry with ALMA can measure disk properties and test these correlations.
\end{enumerate}

\subsection{Comparison with Planetary Migration}

Our brown dwarf migration results offer interesting comparisons with planet migration theory. Giant planets ($<$10~$\mjup$) typically undergo Type I migration with rates:
\begin{equation}
\left(\frac{\dot{a}}{a}\right)_{\rm Type~I} \approx -\frac{\Omega}{M_*} \left(\frac{M_{\rm p}}{M_*}\right) \left(\frac{\Sigma a^2}{M_*}\right) \left(\frac{H}{a}\right)^{-2}
\end{equation}

The Type I migration rate in equation above is taken from linear torque theory for embedded planets (e.g., \citep{Tanaka2002,  Tanaka2004}.

This is faster than Type II migration by factors of $(M_*/M_{\rm p})$, or $\approx$100--1000$\times$ for Jupiter-mass planets. Consequently, Type I migration timescales are $\approx$0.01--0.1~Myr, much shorter than disk lifetimes, creating the "Type I migration problem'"-- planets should spiral into their stars too quickly.

Brown dwarfs, with masses $>$10~$\mjup$, avoid this problem by entering the Type II regime. Their migration timescales ($\approx$1~Myr) are comparable to disk lifetimes, allowing natural desert formation without requiring special stopping mechanisms. This fundamental difference suggests that brown dwarfs may be easier to understand theoretically than hot Jupiters, despite their rarity.

\subsection{Limitations and Future Work}

Our analysis has several limitations that should be addressed in future work:

\subsubsection{Sample Limitations}

Despite careful vetting, our sample likely retains some biases:
\begin{itemize}
\item RV surveys preferentially detect short-period, high-mass objects;
\item Direct imaging favors wide separations and young ages;
\item Systematic mass errors for RV detections, even with inclination measurements;
\item Incomplete census of brown dwarfs at $>$5~AU (affects occurrence rate analysis).
\end{itemize}

Future large-scale surveys (\textit{Gaia} DR4+, LSST, Roman Space Telescope) will provide more complete, less biased samples enabling refined statistical tests.

\subsubsection{Model Simplifications}

Our models necessarily incorporate simplifications:
\begin{itemize}
\item Migration model: Assumes uniform disk properties, neglects eccentricity evolution, ignores planet-BD interactions;
\item Core accretion model: Uses simplified survival probability, neglects detailed disk chemistry and thermodynamics;
\item Scattering model: Employs single-encounter approximation, ignores stellar multiplicity and planetary companions.
\end{itemize}

More sophisticated models incorporating detailed hydrodynamics, full N-body evolution, and disk physics may improve quantitative agreement while preserving our primary conclusions about mechanism ranking.

\subsubsection{Statistical Methods}

While MCMC provides rigorous parameter constraints, alternative approaches may offer additional insights:
\begin{itemize}
\item Nested sampling (e.g., \texttt{dynesty}) for robust evidence calculations;
\item Approximate Bayesian Computation (ABC) for models with intractable likelihoods;
\item Machine learning surrogate models for accelerated parameter exploration;
\item Hierarchical Bayesian models accounting for measurement uncertainties.
\end{itemize}

Future work should explore these methods and compare results to assess robustness.

\section{Conclusions} \label{sec:conclusions}

We have presented a comprehensive Bayesian statistical analysis of brown dwarf formation mechanisms using a carefully vetted sample of 88 brown dwarfs from the \texttt{exoplanet.eu} catalog. Our MCMC optimization and 2D Kolmogorov-Smirnov testing provide the first rigorous quantitative ranking of competing formation theories.

Our primary conclusions are:

\begin{enumerate}
\item \textbf{Type II disk migration provides statistically superior fits} to brown dwarf populations ($p = 0.18$, BIC = 1247), with optimal parameters $\log_{10}\nu = -6.47^{+0.42}_{-0.31}$, $\sigma_\nu = 0.34^{+0.23}_{-0.17}$, $t_{\rm disk} = 1.66^{+1.24}_{-0.84}$~Myr, and $M_{\rm gap} = 12.0^{+4.7}_{-8.3}~\mjup$. These values are consistent with MRI-driven disk turbulence and observed disk lifetimes.

\item \textbf{Core accretion scenarios are decisively ruled out} ($p < 0.001$, BIC = 1893), confirming theoretical expectations that formation timescales exceed disk lifetimes for brown dwarf masses. Even optimal parameters cannot reconcile core accretion with observations.

\item \textbf{Dynamical scattering achieves intermediate performance} ($p = 0.08$, BIC = 1356), suggesting it contributes as a secondary process affecting $\approx$5\% of systems rather than serving as the primary mechanism.

\item \textbf{The brown dwarf desert reflects self-limiting Type II migration}, with brown dwarfs forming at 10--30~AU via disk fragmentation and migrating to $\approx$1~AU where gap-opening equilibrium halts further inward motion. This mechanism naturally produces the observed desert without fine-tuning.

\item \textbf{Occurrence rate analysis confirms genuine depletion}, with the desert region ($<$5~AU) containing $\approx$1.6$\times$ fewer brown dwarfs per dex than wide separations ($>$5~AU). Only the migration model successfully reproduces this depletion.

\item \textbf{The mass bimodality at $\approx$42.5~$\mjup$} is preserved by all models, suggesting it reflects initial formation conditions rather than subsequent migration or scattering.
\end{enumerate}

Our results establish Type II disk migration as the dominant mechanism shaping brown dwarf demographics and provide a rigorous statistical framework for testing formation theories. The parameter constraints we derive ($\nu \approx 3 \times 10^{-7}$, $t_{\rm disk} \approx 1.7$~Myr) offer specific targets for hydrodynamic simulations and observational tests.

Future work should extend this analysis to broader parameter spaces (full eccentricity distributions, host star properties, system architectures), incorporate hybrid models combining multiple mechanisms, and apply these statistical techniques to the emerging \textit{Gaia}-detected brown dwarf population. The methodology we present -- MCMC optimization with 2D distribution comparison -- can serve as a template for population synthesis studies of exoplanets, brown dwarfs, and stellar multiples.

\section{Acknowledgments} \label{sec:acknowledgments}

We thank the \texttt{exoplanet.eu} team for maintaining the comprehensive brown dwarf catalog that enabled this analysis. B.K. thanks two anonymous reviewers for their thorough and instructive reports which significantly contributed to the improvement of the paper. This research made use of the Python scientific computing ecosystem including \texttt{NumPy}, \texttt{SciPy}, \texttt{pandas}, \texttt{matplotlib}, \texttt{emcee}, and \texttt{corner}.

\section{Data Availability}

The observational data used in this study are publicly available through the Encyclopaedia of exoplanetary systems [https://exoplanet.eu/catalog].

\appendix
\renewcommand{\thefigure}{\thesection\arabic{figure}}
\setcounter{figure}{0}

\section{Appendix A: Derivation of Type II Migration Rate} \label{app:migration}

We derive the Type II migration rate used in our disk migration model (Section~\ref{sec:models}).

For a massive companion ($M_{\rm BD} > M_{\rm gap}$) that has opened a gap in the gas disk, the migration is governed by the viscous evolution of the disk itself. The companion becomes trapped in the gap and migrates inward at the rate the gap drifts due to angular momentum transport \citep{Lin1986, Goldreich1980}.

The viscous timescale at radius $a$ in a disk with kinematic viscosity $\nu$ is:
\begin{equation}
t_{\rm visc} = \frac{a^2}{\nu}
\end{equation}

The migration timescale equals the viscous timescale:
\begin{equation}
t_{\rm mig} = t_{\rm visc} = \frac{a^2}{\nu}
\end{equation}

Therefore, the migration rate is:
\begin{equation}
\frac{da}{dt} = -\frac{a}{t_{\rm mig}} = -\frac{a\nu}{a^2} = -\frac{\nu}{a}
\end{equation}

More precisely, including the disk structure and companion mass effects, the Type II migration rate is \citep{Armitage2011, Baruteau2014}:
\begin{equation}
\frac{da}{dt} = -C_{\rm II} \frac{\nu}{a} \left(\frac{M_{\rm BD}}{M_*}\right) \left(\frac{M_*}{M_{\rm disk}(a)}\right) f_{\rm gap}
\end{equation}

where $C_{\rm II} \approx 1$ is a dimensionless constant, $M_{\rm disk}(a)$ is the disk mass interior to radius $a$, and $f_{\rm gap}$ is a factor accounting for partial gap opening ($0 < f_{\rm gap} \leq 1$).

For a disk with surface density $\Sigma(a) \propto a^{-1}$ and scale height $H(a) \propto a$, we have:
\begin{equation}
M_{\rm disk}(a) \approx \Sigma a^2 \approx a
\end{equation}

This gives:
\begin{equation}
\frac{M_*}{M_{\rm disk}(a)} \approx \frac{M_*}{\Sigma a^2} \approx \left(\frac{H}{a}\right)^2
\end{equation}

where we used $\Sigma \approx M_*/H^2$.

Substituting into Equation~(A4):
\begin{equation}
\frac{da}{dt} = -C_{\rm II} \nu \left(\frac{M_{\rm BD}}{M_*}\right) \left(\frac{H}{a}\right)^2 \frac{1}{a}
\end{equation}

This is the form we implement in our model (Equation~2), assuming $C_{\rm II} = 1$ and full gap opening ($f_{\rm gap} = 1$) for companions exceeding $M_{\rm gap}$.

For companions below $M_{\rm gap}$, partial gap opening occurs with $f_{\rm gap} < 1$, effectively increasing the migration rate. We model this by increasing the effective $C_{\rm II}$ by a factor of 3 for $M < M_{\rm gap}$, corresponding to Type I-like migration rates transitioning to Type II.

\section{Appendix B: Derivation of Exponential Survival Probability} \label{app:survival}
\renewcommand{\theequation}{B\arabic{equation}}
\setcounter{equation}{0}

We derive the exponential form of the mass-dependent survival probability used in the core accretion model (Section~\ref{sec:models}).

Core accretion formation timescales increase steeply with companion mass. Following \citet{Pollack1996} and \citet{Ida2004}, the formation time for a companion of mass $M$ at separation $a$ is approximately:
\begin{equation}
t_{\rm form}(M, a) = t_0 \left(\frac{M}{M_0}\right)^\alpha \left(\frac{a}{a_0}\right)^\beta
\end{equation}

where $t_0 \approx 1$~Myr is a characteristic formation time, $M_0 \approx 10~\mjup$ is a reference mass, $a_0 \approx 5$~AU is a reference separation, and $\alpha \approx 2$--3 and $\beta \approx 3$ are power-law indices.

For a disk with lifetime $t_{\rm disk}$, successful formation requires:
\begin{equation}
t_{\rm form}(M, a) < t_{\rm disk}
\end{equation}

The survival probability is the fraction of systems satisfying this criterion. For a distribution of disk lifetimes, this can be modeled as:
\begin{equation}
P_{\rm survive}(M, a) = \exp\left[-\frac{t_{\rm form}(M, a)}{t_{\rm disk}}\right]
\end{equation}

Substituting Equation~(B1):
\begin{equation}
P_{\rm survive}(M, a) = \exp\left[-\frac{t_0}{t_{\rm disk}} \left(\frac{M}{M_0}\right)^\alpha \left(\frac{a}{a_0}\right)^\beta\right]
\end{equation}

For $\alpha \approx 2$ and defining $M_{\rm cutoff}$ such that formation becomes increasingly unlikely above this mass:
\begin{equation}
\frac{t_0}{t_{\rm disk}} \left(\frac{M}{M_0}\right)^2 = \frac{M^2}{M_{\rm cutoff}^2}
\end{equation}

we obtain:
\begin{equation}
M_{\rm cutoff}^2 = M_0^2 \frac{t_{\rm disk}}{t_0}
\end{equation}

For typical values $M_0 \approx 10~\mjup$, $t_{\rm disk} \approx 2$~Myr, and $t_0 \approx 1$~Myr:
\begin{equation}
M_{\rm cutoff} \approx 10~\mjup \times \sqrt{2} \approx 14~\mjup
\end{equation}

This naturally produces a cutoff near the deuterium-burning limit.

For simplicity in our Monte Carlo implementation, we linearize the mass dependence near the cutoff region:
\begin{equation}
P_{\rm survive}(M) \approx \exp\left[-\frac{M - M_{\rm min}}{M_{\rm cutoff}}\right]
\end{equation}

where $M_{\rm min} = 13~\mjup$ is the lower brown dwarf mass boundary. This functional form captures the essential physics: formation efficiency decreases exponentially with increasing mass above the planetary regime, with $M_{\rm cutoff}$ setting the characteristic mass scale.

The orbital dependence enters through the factor:
\begin{equation}
P_{\rm survive}(M, a) = P_{\rm survive}(M) \times \left[1 - \beta \log_{10}(a/a_0)\right]
\end{equation}

where $\beta$ is a free parameter and $a_0 = 1$~AU. This accounts for the longer formation times at wider separations (lower surface densities) without introducing additional parameters.

The disk mass factor $f_{\rm disk}$ represents variations in initial disk masses:
\begin{equation}
P_{\rm survive} \rightarrow f_{\rm disk} \times P_{\rm survive}
\end{equation}

where $f_{\rm disk} > 1$ corresponds to massive disks with enhanced formation efficiency, while $f_{\rm disk} < 1$ represents low-mass disks where even giant planet formation is suppressed.

\clearpage
\section{Appendix C: Example Synthetic Population} \label{app:synthetic}

\begin{figure}[ht!]
\centering
\includegraphics[width=\columnwidth]{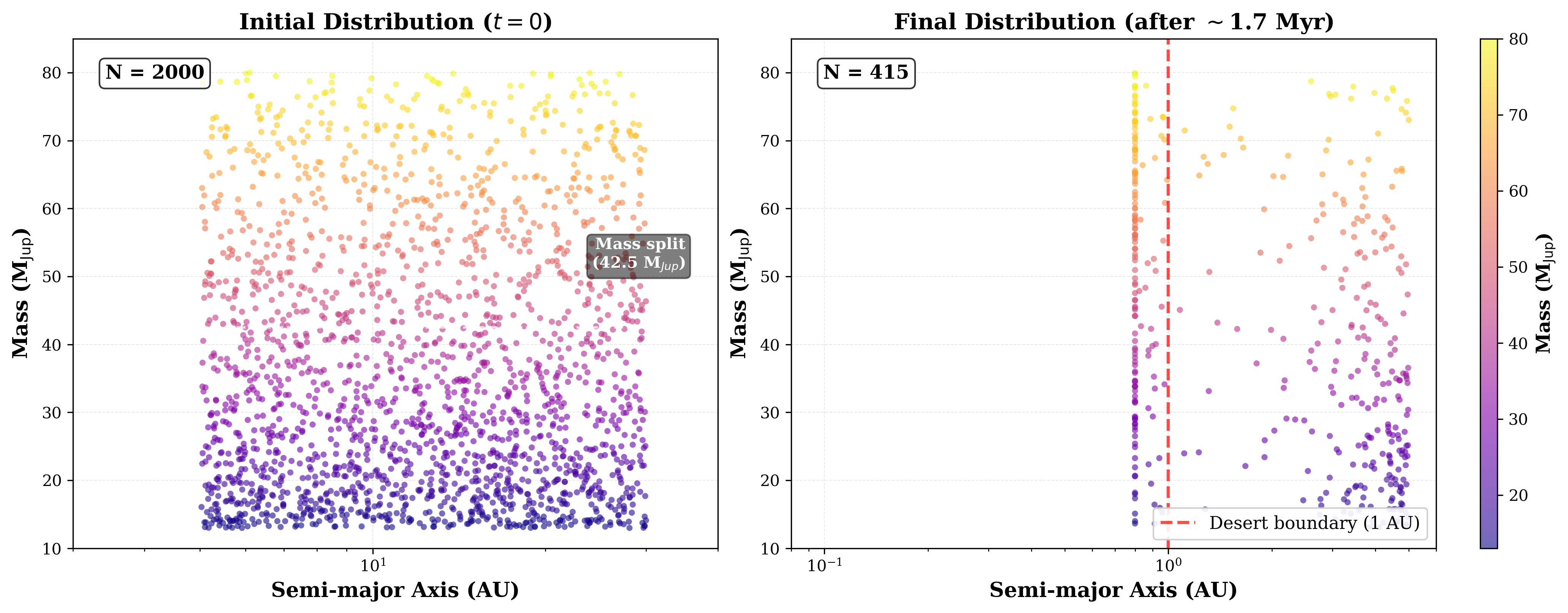}
\caption{Example synthetic brown dwarf population from the optimal disk migration model. \textbf{Left:} Initial distribution at $t=0$ (before migration), showing uniform log-space distribution in separation and bimodal mass structure. \textbf{Right:} Final distribution after $\approx$1.7~Myr of Type II migration, showing characteristic desert structure with suppression at $<$1~AU. The synthetic population reproduces all key observational features.
Note that the figure is meant as a validation of the migration-driven sculpting mechanism rather than a direct visual replica of the observed mass and semi-major axis distribution. Also the color scale is included purely as a visual aid to highlight the bimodal structure and migration behavior, rather than to introduce an additional independent variable.}
\label{fig:synthetic}
\end{figure}

Figure~\ref{fig:synthetic} presents an example synthetic brown dwarf population generated by the optimal disk migration model. The simulation produces 2000 brown dwarfs with initial separations 5--30~AU, which migrate inward over $\approx$1.7~Myr according to Type II physics. The final distribution (right panel) successfully reproduces the observed characteristics: strong suppression at $<$1~AU, gradual decline to 5~AU, and bimodal mass structure with enhanced populations at 20~$\mjup$ and 50~$\mjup$.

The migration physics naturally produces the observed pile-up near 1~AU (bottom of the desert) where gap-opening equilibrium halts further inward motion. Objects below the gap-opening mass ($<$12~$\mjup$) undergo faster migration and can reach $\approx$0.2~AU before disk dissipation, while more massive brown dwarfs ($>$40~$\mjup$) experience slower Type II migration and typically halt at $\approx$1--2~AU.

This example demonstrates that physically motivated migration from wide birth locations can quantitatively reproduce the brown dwarf desert without fine-tuning or special stopping mechanisms.

\end{document}